\documentstyle [epsfig, 12pt]{article}
\begin{document}
\bigskip

\bigskip

\begin{center}
{\Large {\bf CCD BVRI and 2MASS Photometry of the Poorly Studied Open Cluster NGC 6631}}
\end{center}

\bigskip
\begin{center}
{\bf Tadross, A. L.$^a$, Bendary, R.$^a$, Priya, H.$^b$, Essam, A.$^a$, Osman, A.$^a$}
\bigskip
\\ {\it $^a$ National Research Institute of Astronomy and Geophysics, 11421 - Helwan, Cairo, Egypt.
\\ $^b$ Muffakham Jah College of Engineering and Technology, Banjara Hills, Hyderabad 500 034, India.
\\E-mail: altadross@yahoo.com}
\end{center}
\bigskip
{\bf Abstract.} Here we have obtained the {\it BVRI CCD} photometry down to a limiting magnitude of $V \sim$ 20 for the southern poorly studied open cluster NGC 6631. It is observed from the {\it 1.88 m} Telescope of Kottamia Observatory in Egypt. About 3300 stars have been observed in an area of $\sim 10^{\prime} \times 10^{\prime}$ around the cluster center. The main photometric parameters have been estimated and compared with the results that determined for the cluster using {\it JHKs 2MASS} photometric database. The cluster's diameter is estimated to be 10 arcmin; the reddening E(B-V)= 0.68 $\pm$ 0.10 mag, E(J-H)= 0.21 $\pm$ 0.10 mag, the true modulus (m-M)$_{o}$= 12.16 $\pm$ 0.10 mag, which corresponds to a distance of 2700 $\pm$125 pc and age of 500 $\pm$ 50 Myr.
\bigskip

Open star clusters (OCs) are ideal objects for studying the main properties of the Milky Way Galaxy, i.e. star formation, stellar evolution, and distance scale of the Galaxy. The fundamental parameters of an open cluster; e.g. distance, age, and interstellar extinction; can be determined by comparing color-–magnitude (CM) and color–-color (CC) diagrams with the modern theoretical models. As we know, more than half of the currently catalogued open clusters ($\sim$ 2000 OCs) have been poorly studied or even unstudied at all. The current paper is thus a part of our goal to obtain good-quality photometric data for poorly studied or unstudied OCs and to estimate their fundamental parameters more accurately.

 The open cluster NGC 6631 located in the direction of Galactic center at constellation SCT-Scutum at 2000.0 coordinates $\alpha=18^{h} \ 27^{m} \ 11^{s}, \ \delta=-12^{\circ} \ 01^{'} \ 48^{''}, \ \ell= 19.47^{\circ}, \ b= -0.19^{\circ}$.

 The cluster is classified as II2m by Ruprecht (1966) and as II1m by Lyng{\aa} (1987). Alter et al. (1970) has presented some rough estimations for angular diameter of this cluster ranging from 4 to 16 arcmin, while the distance varies from 880 to 5000 pc. We can say that no real study has been done for this cluster before Ram Sagar et al. (2001). They presented the first {\it VI CCD} photometric study of NGC 6631 and estimated the cluster main parameters by getting the best fit of the theoretical isochrones on V$\sim$(V--I) diagram of the cluster. They obtained distance = 2.6 $\pm$ 0.5 kpc, age = 400 $\pm$ 100 Myr, and E(V-I) = 0.60 $\pm$ 0.05 mag. The cluster diameter estimated from the radial density profile was 4.8 $\pm$ 0.5 pc, while the mass function of the cluster has a slope of 2.1 $\pm$ 0.5.

This paper is organized as follows: Observations and data reduction is presented in section 2. Radial Density Profile of the cluster is described in Sections 3. Analysis of the color-magnitude diagrams are declared in Section 4. The mass function of the cluster is estimated and discussed in Section 5. Finally, the conclusion of our study is devoted to section 6.

\section{Observations and Data Reductions}

The {\it BVRI CCD} photometric observations of the star cluster NGC 6631 were obtained during June 12 to 14, 2012, using the Newtonian focus scale 22.53 arcsec per mm of the 1.88 m Reflector Telescope at Kottamia Observatory in Egypt. Fig. 1 represents the image of the cluster in {\it I} band.
The characteristics of the CCD camera are listed in Table (1), while the observation log is given in Table (2). The images were bias subtracted and flat-fielded using standard procedures in IRAF and the photometry was done using IRAF/DAOPHOT (Stetson, 1987, 1992). The standard stars field SA 110 230 (Landolt, 1992) were observed for standardization and the APPHOT photometry was used to derive the observed magnitudes. Extinction coefficients and zero points were determined to standardize the data. Fig. 2 shows the magnitude versus error for the BVRI bands from the DAOPHOT photometry. The total number of stars in the B, V, R and I bands are 1259, 1806, 3043 and 3250 respectively. The limits of errors in our final photometry are 0.363, 0.068, 0.034 and 0.094 respectively.

On the other hand, the near-{\it IR JHK} data are taken from the digital Two Micron All Sky Survey (2MASS) of Skrutskie et al. (2006). It is uniformly scanning the entire Sky in three {\it IR} bands J (1.25 $\mu$m), H (1.65 $\mu$m) and Ks (2.17 $\mu$m). Data extraction has been performed at a preliminary radius of 10 arcmin using the known tool of VizieR for {\it 2MASS}\footnote{\it http://vizier.cfa.harvard.edu/viz-bin/VizieR?-source=II/246} database. A cutoff of photometric completeness limit at $J \geq 16.5$ mag is applied to the data to avoid the over-sampling (cf. Bonatto et al. 2004). Also, for photometric quality, stars with errors in J, H and Ks bigger than 0.20 mag have been excluded (cf. Tadross 2011 and references therein).

\begin{figure}
\begin{center}
      {\includegraphics[width=6cm]{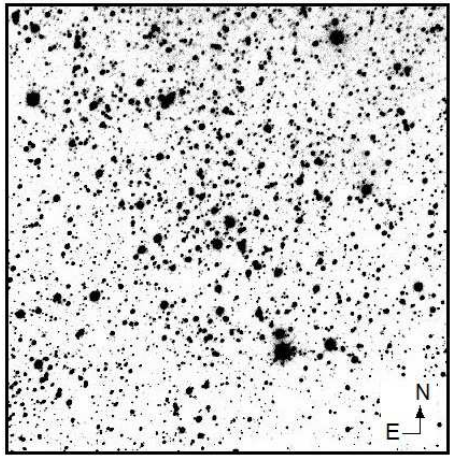}}
      \end{center}
      \caption{The {\it CCD I} image of open star cluster NGC 6631 as observed by 1.88 m Kottamia Telescope of Egypt. North is up, East on the left.}
\end{figure}
\begin{figure}
\begin{center}
      {\includegraphics[width=8cm]{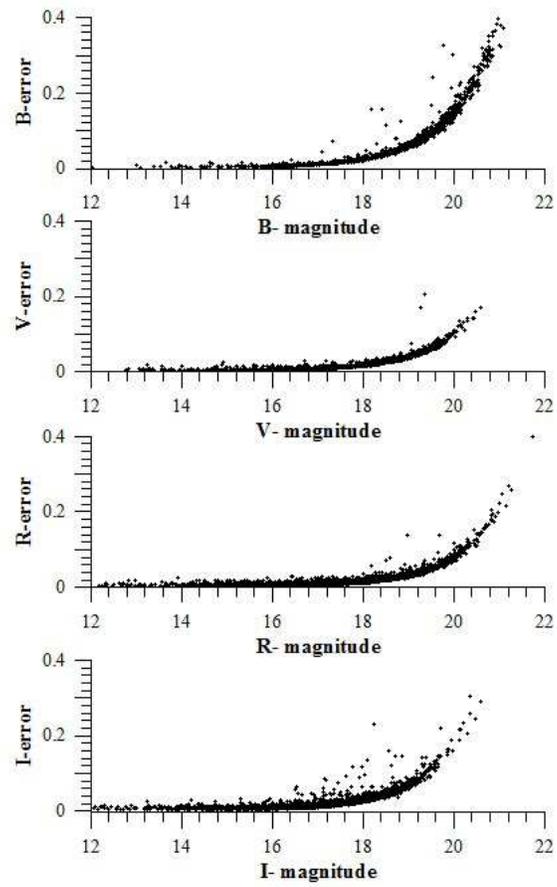}}
      \end{center}
      \caption{The {\it BVRI} errors of the observed magnitudes for the stars of NGC 6631.}
\end{figure}

\begin{table*}
\renewcommand{\arraystretch}{}
\caption[]{The characteristics of CCD camera used in the observations.}
\begin{minipage}{\textwidth}
\begin{tabular}{ll}
 \\
\hline
Type                  &  EEV CCD 42-40 \\
Version               &  Back-illuminated with BPBC (Basic Process Broadband Coating) \\
Format                &   2048 $\times$ 2048 pixel$^{2}$ \\
Pixel size            &  13.5 $\times$ 13.5 µm$^{2}$ \\
Grade                 & 0 \\
Dynamic range         &  30:1 \\
A/D converter        &  16 bit \\
Imaging area          & 27.6 $\times$ 27.6 mm$^{2}$ \\
Read out noise@20 KHz & 3.9 $e^{-}/$ pixel \\
Gain                  &  2.26 $e^{-}/ADU$ (Left amplifier) and 2.24 $e^{-}/ADU$ (Right amplifier) \\
\hline \label{OCdata}
\end{tabular}
\end{minipage}
\end{table*}
\begin{table*}
\renewcommand{\arraystretch}{}
\caption[]{The Log of Optical Photometric CCD Observations.}
\begin{minipage}{\textwidth}
\begin{tabular}{lll} \hline
Filter & No. of frames & Exposure time in seconds\\
\hline
B & 8 & 60 \\
  & 3 & 80 \\
  & 4 & 120 \\
V & 4 & 120 \\
R & 8 & 120 \\
I & 4 & 120 \\
\hline \label{OCdata}
\end{tabular}
\end{minipage}
\end{table*}

\section{The Radial Density Profile of the cluster}

To establish the radial density profile (RDP) of NGC 6631, we counted the stars of the cluster (taken from 2MASS database) within concentric shells in equal incremental steps of 0.1 arcmin from the cluster center. We repeated this process for $0.1<r\leq0.2$ up to 10 arcmin, i.e. the stellar density is derived out to the preliminary radius of the cluster. The stars of the next steps should be subtracted from the later ones, so that we obtained only the amount of the stars within the relevant shell's area, not a cumulative count. Finally, we divided the star counts in each shell to the area of that shell those stars belong to. The density uncertainties in each shell were calculated using Poisson noise statistics. Fig. 3 shows the RDP for NGC 6631 to the maximum angular separation of 5 arcmin where the decay becomes asymptotically at that point. Applying the empirical King model (1966), where it parameterizes the density function $\rho(r)$ as:

\begin{center}
{\Large $\rho(r)=f_{bg}+\frac{f_{0}}{1+(r/r_{c})^{2}}$}
\end{center}

where $f_{bg}$, $f_{0}$ and $r_{c}$ are background, central star density and the core radius of the cluster respectively. In this context, $f_{bg} \sim$ 36 stars per arcmin$^{2}$, $f_{0}$ = 42 stars per arcmin$^{2}$, and $r_{c}$ = 0.59 arcmin. According to the next section, consequently, the radial diameter of the cluster is determined to be 7.85 pc.

\begin{figure}
\begin{center}
      {\includegraphics[width=8.5cm]{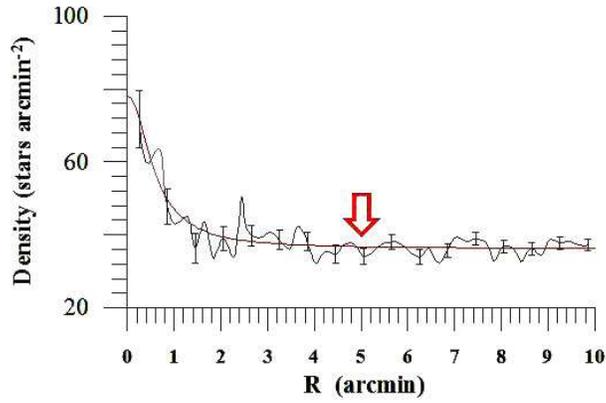}}
      \end{center}
      \caption{The radial density distribution of the stars in NGC 6631. The decay of the density reaches a value of $\rho=37$ stars/arcmin$^{2}$ at 5.0 arcmin, where the decay becomes asymptotic. The curved solid line represents the fitting of King (1966) model. Error bars are determined from sampling statistics ($1/\sqrt{N}$ where N is the number of stars used in the density estimation at that point). The background field density $f_{bg} \sim$ 36 stars per arcmin$^{2}$. The core radius $r_{c}$ = 0.59 arcmin.}
\end{figure}

\section{Color-Magnitude Diagrams}

The Color-Magnitude Diagrams {\it (CMDs)} of the observed stars: V$\sim$(B--V), V$\sim$(V--I), V$\sim$(V--R), R$\sim$(R--I), and of the obtained {\it JHK-2MASS}: J$\sim$(J--H) and K$\sim$(J--K) are constructed for the cluster. The theoretical isochrones of Padova\footnote{http://stev.oapd.inaf.it/cgi-bin/cmd} that computed by Marigo et al. (2008) are used in fitting processes. Several solar isochrones (Z $\sim$ 0.02) of different ages have been applied to the {\it  CMDs} of NGC 6631. The best fittings for {\it BVRI} diagrams are obtained at distance modulus of 14.20 $\pm$ 0.10 mag, age of 500 $\pm$ 50 Myr, and reddening of 0.68, 1.00, 0.54, and 0.47 $\pm$ 0.10 mag respectively, from left to right as shown in Fig. 4. On the other hand, the fittings for {\it JHK-2MASS} diagrams are obtained at distance modulus of 12.75 $\pm$ 0.10 mag, age of 500 $\pm$ 50 Myr, and reddening of 0.21 and 0.33 $\pm$ 0.10 mag, from left to right as shown in Fig. 5. The resulting total visual absorption is taken from the ratio $A_{v}/E(B-V)$ = 3.1, following Garcia et al. (1988).

{\it JHK-2MASS} data has been corrected for interstellar reddening using the coefficient ratios $\frac {A_{J}}{A_{V}}=0.276$ and $\frac {A_{H}}{A_{V}}=0.176$, which were derived from absorption rations in Schlegel et al. (1998), while the ratio $\frac {A_{K_s}}{A_{V}}=0.118$ was derived from Dutra et al. (2002). Applying the calculations of Fiorucci \& Munari (2003) for the color excess of {\it 2MASS} photometric system; we ended up with the following results: $\frac {E_{J-H}}{E_{B-V}}=0.309\pm0.130$, $\frac {E_{J-K_s}}{E_{B-V}}=0.485\pm0.150$, where R$_{V}=\frac {A_{V}}{E_{B-V}}= 3.1$. Also, we can de-reddened the distance modulus using these formulae:  $\frac {A_{J}}{E_{B-V}}$= 0.887, $\frac {A_{K_s}}{E_{B-V}}$= 0.322. Therefore, the true distance modulus is calculated to be $(V-M_{v})_{o} =12.16 \pm 0.10$\,mag, corresponding to a distance of $2700\pm 125$\,pc.

After estimating the cluster's distance from the Sun, $R_{\odot}$, the distance from the galactic center ($R_{g}$), the projected distances on the galactic plane from the Sun ($X_{\odot}~\&~Y_{\odot}$) and the distance from the galactic plane ($Z_{\odot}$) are estimated to be 6000, --2545, 900 and --8.95\,pc respectively.

\begin{figure}
\begin{center}
      {\includegraphics[width=14cm]{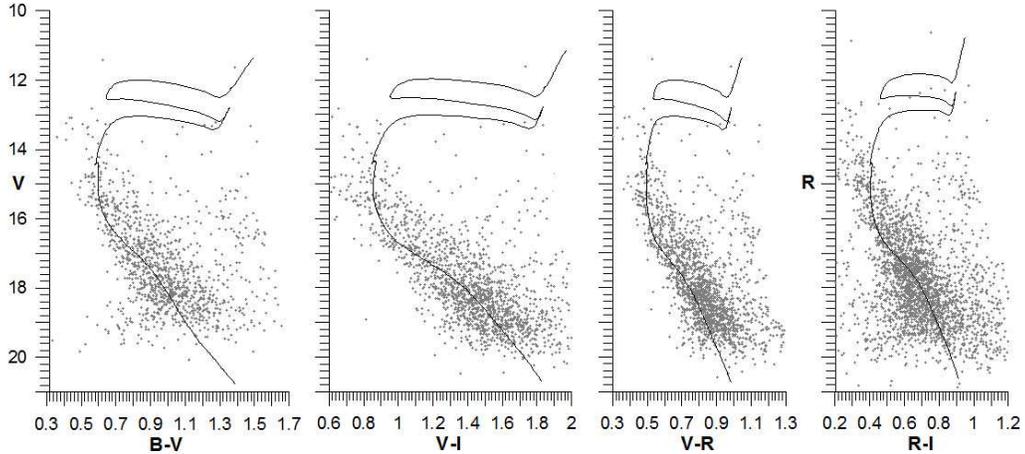}}
      \end{center}
      \caption{Theoretical {\it BVRI}-isochrones fit to the observed {\it CMDs} of NGC 6631. The distance modulus is found to be 14.20 mag, and the color excesses are found to be (from left to right) 0.68, 1.00, 0.54 and 0.47 mag respectively.}
\end{figure}
\begin{figure}
\begin{center}
      {\includegraphics[width=8cm]{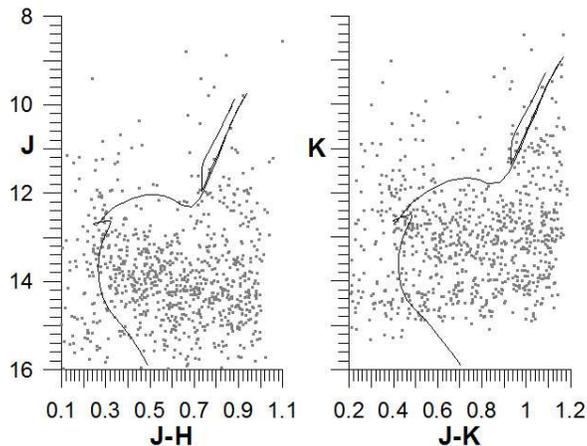}}
      \end{center}
      \caption{Theoretical {\it JHK}-isochrones fit to the obtained {\it CMDs} of NGC 6631. The distance modulus is found to be 12.75 mag, and the color excesses are found to be (from left to right) 0.21 and 0.33 mag respectively.}
\end{figure}

\section{The Mass Function of NGC 6631}

It is difficult to determine the membership of the cluster using only the stellar RDP. The stellar membership is found more precisely for those stars are closer to the cluster's center and in the same time very near to the main-sequence (MS) in CMDs. These MS stars are very important in determining the luminosity and mass functions of the investigated cluster.

The number of stars per luminosity interval, or in other words, the number of stars in each magnitude bin, gives us what so-called the luminosity function (LF) of the cluster. In order to estimate the LF of NGC 6631, we count the observed stars in terms of absolute magnitude after applying the distance modulus. The magnitude bin intervals are selected to include a reasonable number of stars in each bin and for the best possible statistics of the luminosity and mass functions. From LF, we can infer that the massive bright stars seem to be centrally concentrated more than the low masses and fainter ones (Montgomery et al. 1993).

The LF and the mass function (MF) are correlated to each other according the known Mass-luminosity relation. The accurate determination of both of them (LF \& MF) suffers from some problems e.g. the contamination of field stars; the observed incompleteness at low-luminosity (or low-mass) stars; and mass segregation, which may affect even poorly populated, relatively young clusters (Scalo 1998). On the other hand, the properties and evolution of a star are closely related to its mass, so the determination of the initial mass function (IMF) is needed, that is an important diagnostic tool for studying large quantities of star clusters. IMF is an empirical relation that describes the mass distribution (a histogram of stellar masses) of a population of stars in terms of their theoretical initial mass (the mass they were formed with). The IMF is defined in terms of a power law as follows:

\begin{center}
{\Large $\frac{dN}{dM} \propto M^{-\alpha}$}
\end{center}

where $\frac{dN}{dM}$ is the number of stars of mass interval (M:M+dM), and $\alpha$ is a dimensionless exponent. The IMF for massive stars ($>$ 1 $M_{\odot}$) has been studied and well established by Salpeter (1955), where $\alpha$ = 2.35. This form of Salpeter shows that the number of stars in each mass range decreases rapidly with increasing mass. Fig. 6 shows that the {\it BVRI} and {\it JHK} mass functions of NGC 6631, where the slopes of the two MFs close to Salpeter's value. The right panel of Fig. 6 seems to complete the left panel. The mean slope of the mass function taken to be 2.3$\pm$ 0.05.

\begin{figure}
\begin{center}
      {\includegraphics[width=12.5cm]{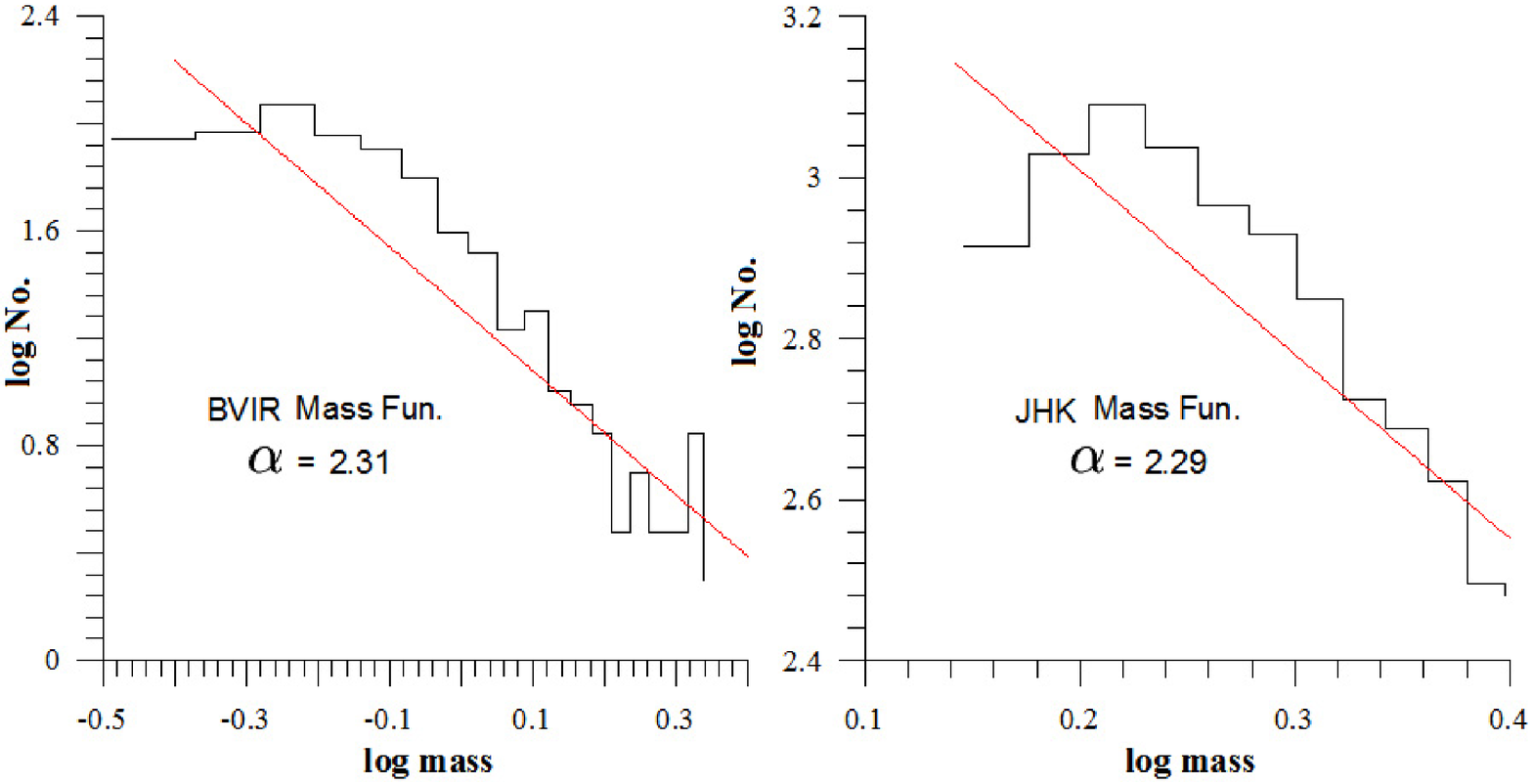}}
      \end{center}
      \caption{The {\it BVRI} and {\it JHK} mass functions of NGC 6631. The slopes of the two panels are close to Salpeter's value, see Sec. 5.}
\end{figure}

\section{Conclusion}

 The open star cluster NGC 6631 has been observed using {\it BVRI} pass-band of the {\it 1.88 m} Kottamia Telescope of Egypt. The main astrophysical properties of the cluster have been estimated and confirmed by the {\it JHK} 2MASS bass-band data. It is noted that the determination of the cluster radius made by the uniformity of 2MASS database allow us to obtain reliable data on the projected distribution of stars for large extensions to the clusters' halos. However, a comparison between the results of the present work with those of Ram Sagar (2001) is given in Table 3.

 \begin{table}
\caption{Comparison between the present and previous studies.}
\begin{tabular}{lll}
\hline\noalign{\smallskip}Parameter&The present work&Ram Sagar (2001)
\\\hline\noalign{\smallskip}
E(V-I)&1.00 $\pm$ 0.10 mag.& 0.60 $\pm$ 0.05 mag.\\
E(B-V)&0.68 $\pm$ 0.10 mag.& ---\\
E(V-R)&0.54 $\pm$ 0.10 mag.& ---\\
E(J-H)&0.21 $\pm$ 0.10 mag.& ---\\
E(J-K)&0.33 $\pm$ 0.10 mag.& ---\\
(m-M)$_{BVRI}$&14.20 $\pm$ 0.10 mag.&13.50 $\pm$ 0.30 mag.\\
(m-M)$_{JHK}$&12.75 $\pm$ 0.10 mag.& --- \\
Distance ($R_{\odot}$)&2700 $\pm$125 pc.&2600 $\pm$ 500 pc.\\
$R_g$&6000 pc&---\\
$X_{\odot}$&--2545 pc&---\\
$Y_{\odot}$&900 pc&---\\
$Z_{\odot}$&--8.95 pc&---\\
Age&500 $\pm$ 50 Myr.&400 $\pm$ 100 Myr.\\
Metallicity {\it (Z)}&0.02&0.05\\
Radius& 5.0 arcmin. & 3.2 arcmin.\\
Linear diameter& 7.85 pc& 4.80 pc\\
Core radius&$\approx$ 0.59 arcmin. ($\sim$ 0.5 pc.)&---\\
Mass function slope& 2.3$\pm$ 0.05& 2.1$\pm$ 0.50\\
\hline{\smallskip}
\end{tabular}
\end{table}
\bigskip
{\bf acknowledgements}\\
This paper is a part of the project No. STDF-1335; funded by Science \& Technology Development Fund (STDF) under the Egyptian Ministry for Scientific Research. The project team expresses their deep appreciation to the administrators of STDF and its organization.
\\
\\
\newpage
{\bf REFERENCES}
\\
Alter, G. et al. 1970, Catalogue of star clusters and associations, 2nd ed., Akademiai Kiado, Budapest \\
Bonatto, Ch., Bica, E., Girardi, L. 2004, A\&A, 415, 571 \\
Dutra, C., Santiago, B., Bica, E. 2002, A\&A, 381, 219 \\
Fiorucci, M., Munari, U. 2003, A\&A, 401, 781 \\
Garcia, B., Clari$\acute{a}$, J., Levato, H. 1988, Ap\&SS, 143, 377 \\
King, I. 1966, AJ, 71, 64 \\
Landolt, A. U. 1992, AJ, 104, 340 \\
Lyng{\aa}, G., Palous, J. 1987, Astro. Astrophys., 188, 35 \\
Marigo, P., et al. 2008, A\&A, 482, 883 \\
Montgomery, K.A., Marschall, L.A., Janes, K.A. 1993, AJ, 106, 181 \\
Ram Sagar, et al. 2001, Bull. Astr. Soc. India, 29, 519 \\
Ruprecht, J. 1966, Bulletin of the Astronomical Institute of Czechoslovakia, 17, 33 \\
Salpeter, E. 1955, ApJ, 121, 161 \\
Scalo, J. 1998, ASPC, 142, 201 \\
Schlegel, D., et al. 1998, ApJ, 500, 525 \\
Skrutskie, M., et al. 2006, AJ, 131, 1163 \\
Stetson, P.B. 1987, PASP, 99, 191 \\
Stetson, P.B. 1992, IAU col. 136, 291 \\
Tadross, A. L. 2011, JKAS, 44, 1 \\

\end{document}